\documentclass[twocolumn,noshowpacs,preprintnumbers,amsmath,amssymb,prl]{revtex4}
\usepackage{graphicx}
\usepackage{multirow}

\begin{document}

\title{Tunable Kondo effect in a single donor atom}

\author{G.P.\,Lansbergen$^{1}$}
\author{G.C.\,Tettamanzi$^{1}$}
\author{J.\,Verduijn$^{1}$}
\author{N.\,Collaert$^{2}$}
\author{S.\,Biesemans$^{2}$}
\author{M.\,Blaauboer$^{1}$}
\author{S.\,Rogge$^{1}$}

\affiliation{$^{1}$Kavli Institute of Nanoscience, Delft University of Technology, Lorentzweg 1, 2628 CJ Delft, The Netherlands}
\affiliation{$^{2}$InterUniversity Microelectronics Center (IMEC), Kapeldreef 75, 3001 Leuven, Belgium}

\begin{abstract}
The Kondo effect has been observed in a single gate-tunable atom. The measurement device consists of a single As dopant incorporated in a Silicon nanostructure. The atomic orbitals of the dopant are tunable by the gate electric field. When they are tuned such that the ground state of the atomic system becomes a (nearly) degenerate superposition of two of the Silicon valleys, an exotic and hitherto unobserved {\it valley} Kondo effect appears. Together with the ``regular'' spin Kondo, the tunable valley Kondo effect allows for reversible electrical control over the symmetry of the Kondo ground state from an SU(2)- to an SU(4) -configuration.
\end{abstract}

\date{\today}

\maketitle
The addition of magnetic impurities to a metal leads to an anomalous increase of their resistance at low temperature. Although discovered in the 1930's, it took until the 1960's before this observation was satisfactorily explained in the context of exchange interaction between the localized spin of the magnetic impurity and the delocalized conduction electrons in the metal \cite{JunKondo}. This so-called Kondo effect is now one of the most widely studied phenomena in condensed-matter physics \cite{Hewson} and plays a mayor role in the field of nanotechnology. Kondo effects on single atoms have first been observed by STM-spectroscopy and were later discovered in a variety of mesoscopic devices ranging from quantum dots and carbon nanotubes to single molecules \cite{WingreenOverview}. 

Kondo effects, however, do not only arise from localized spins: in principle, the role of the electron spin can be replaced by another degree of freedom, for example orbital momentum \cite{Cox}. The simultaneous presence of both a spin- and an orbital degeneracy gives rise to an exotic SU(4)-Kondo effect, where "SU(4)" refers to the symmetry of the corresponding Kondo ground state \cite{Inoshita, Borda}. SU(4) Kondo effects have received quite a lot of theoretical attention \cite{Borda, Zarand}, but so far little experimental work exists \cite{Pablo}.

The atomic orbitals of a gated donor in Si consist of linear combinations of the sixfold degenerate valleys of the Si conduction band. The orbital- (or more specifically valley) -degeneracy of the atomic ground state is tunable by the gate electric field. The valley splitting ranges from $\sim$ 1 meV at high fields (where the electron is pulled towards the gate interface) to being equal to the donors valley-orbit splitting ($\sim$ 10-20 meV) at low fields \cite{Koiller, NP}. This tunability essentially originates from a gate-induced quantum confinement transition \cite{NP}, namely from Coulombic confinement at the donor site to 2D-confinement at the gate interface. 

In this article we study Kondo effects on a novel experimental system, a single donor atom in a Silicon nano-MOSFET. The charge state of this single dopant can be tuned by the gate electrode such that a single electron (spin) is localized on the site. Compared to quantum dots (or artificial atoms) in Silicon \cite{Rokhinson, SanquerKondo, Klein}, gated dopants have a large charging energy compared to the level spacing due to their typically much smaller size. As a result, the orbital degree of freedom of the atom starts to play an important role in the Kondo interaction. As we will argue in this article, at high gate field, where a (near) degeneracy is created, the valley index forms a good quantum number and Valley Kondo \cite{Sean} effects, which have not been observed before, appear. Moreover, the Valley Kondo resonance in a gated donor can be switched on and off by the gate electrode, which provides for an electrically controllable quantum phase transition \cite{Roch} between the regular SU(2) spin- and the SU(4) -Kondo ground states.

\begin{figure}
\includegraphics[width=8cm,clip,trim=2cm 0cm 0cm 0cm]{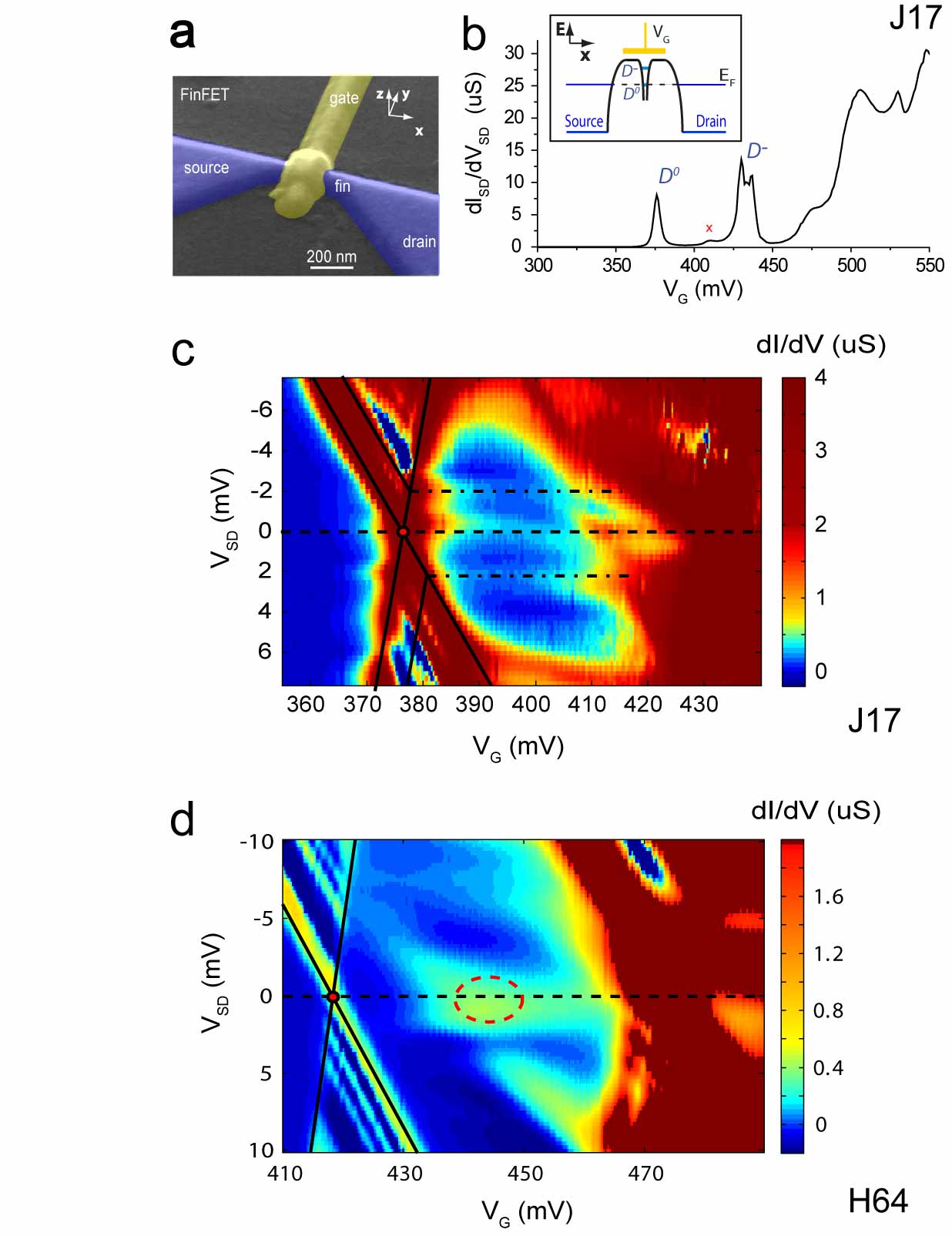}
\caption{Coulomb blocked transport through a single donor in FinFET devices (a) Colored Scanning Electron Micrograph of a typical FinFET device. (b) Differential conductance (dI$_\mathrm{SD}$/dV$_\mathrm{SD}$) versus  gate voltage at V$_\mathrm{SD}=0$. (D$^0$) and (D$^-$) indicate respectively the transport resonances of the one- and two -electron state of a single As donor located in the FinFET channel. Inset: Band diagram of the FinFET along the $x$-axis, with the (D$^0$) charge state on resonance. (c) and (d) Colormap of the differential conductance (dI$_\mathrm{SD}$/dV$_\mathrm{SD}$) as a function of V$_\mathrm{SD}$ and V$_\mathrm{G}$ of samples J17 and H64. The red dots indicate the (D$^0$) resonances and data were taken at 1.6 K. All the features inside the Coulomb diamonds are due to second-order charge fluctuations (see text).} \label{fig1}
\end{figure}

In our experiment we use wrap-around gate (FinFET) devices, see Fig.\,\ref{fig1}(a), with a single Arsenic donor in the channel dominating the sub-threshold transport characteristics \cite{Sellier_PRL}. Several recent experiments have shown that the fingerprint of a single dopant can be identified in low-temperature transport through small CMOS devices \cite{Calvet, Sellier_PRL, Sanquer}. We perform transport spectroscopy (at 4K) on a large ensemble of FinFET devices and select the few that show this fingerprint, which essentially consists of a pair of characteristic transport resonances associated with the one-electron (D$^0$)- and two-electron (D$^-$) -charge states of the single donor \cite{Sellier_PRL}.
From previous research we know that the valley splitting in our FinFET devices is typically on the order of a few meV's. In this Report, we present several such devices that are in addition characterized by strong tunnel coupling to the source/drain contacts which allows for sufficient exchange processes between the metallic contacts and the atom to observe Kondo effects.

Fig.\,\ref{fig1}b shows a zero bias differential conductance (dI$_\mathrm{SD}$/dV$_\mathrm{SD}$) trace at 4.2\,K as a function of gate voltage (V$_\mathrm{G}$) of one of the strongly coupled FinFETs (J17). At the V$_\mathrm{G}$ such that a donor level in the barrier is aligned with the Fermi energy in the source-drain contacts ($E_F$), electrons can tunnel via the level from source to drain (and vice versa) and we observe an increase in the dI$_\mathrm{SD}$/dV$_\mathrm{SD}$. The conductance peaks indicated by (D$^0$) and (D$^-$) are the transport resonances via the one-electron and two-electron charge states respectively. At high gate voltages (V$_G >$ 450\,mV), the conduction band in the channel is pushed below $E_F$ and the FET channel starts to open. The D$^-$ resonance has a peculiar double peak shape which we attribute to capacitive coupling of the D$^-$ state to surrounding As atoms \cite{condmatgrenoble}. The current between the D$^0$ and the D$^-$ charge state is suppressed by Coulomb blockade. 

\begin{figure*}
\includegraphics[width=18cm,clip,trim=0cm 0cm 0cm 0cm]{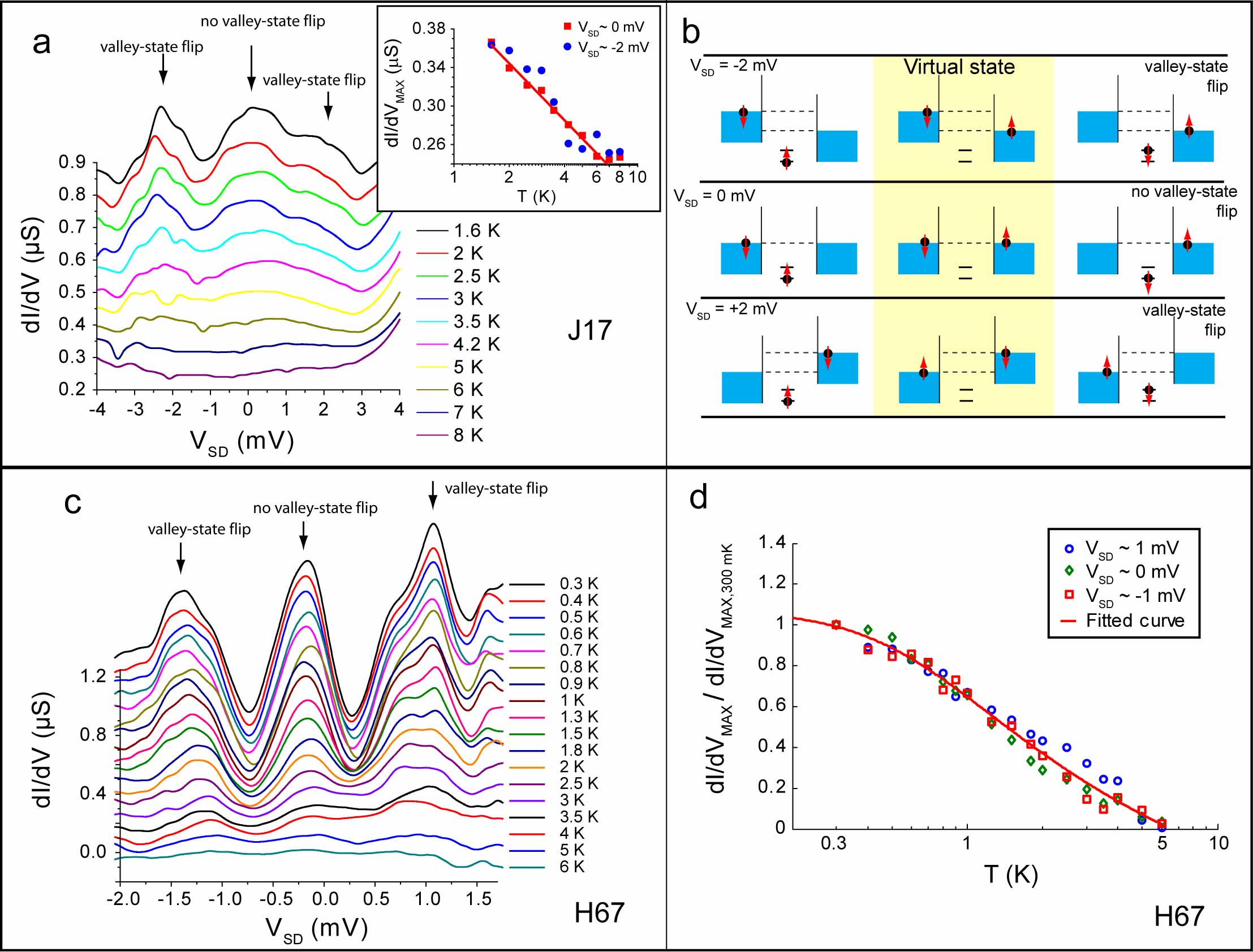}
\caption{Electrical transport through a single donor atom in the Coulomb blocked region (a) Differential conductance 
of sample J17 as a function of V$_\mathrm{SD}$ in the Kondo regime (at V$_\mathrm{G}=395 mV$). For clarity, the temperature traces have been offset 
by 50\,nS with respect to each other. Both the resonances with- and without valley-state flip scale similarly 
with increasing temperature. Inset: Conductance maxima of the resonances at V$_\mathrm{SD}$=-2\,mV and 0\,mV 
as a function of temperature. (b) Schematic depiction of three (out of several) second-order processes underlying 
the zero bias and $\pm \Delta$ resonances. (c) Differential conductance of sample H67 as a function of 
V$_\mathrm{SD}$ in the Kondo regime between 0.3\,K and 6\,K. A linear (and temperature independent) background 
on the order of 1\,$\mu$S was removed and the traces have been offset by 90\,nS with respect to each other for 
clarity. (d) The conductance maxima of the three resonances of (c) normalized to their 0.3\,K value. 
The red line is a fit of the data by Eq.\,\ref{vgl1}.} \label{fig2}
\end{figure*}

The dI$_\mathrm{SD}$/dV$_\mathrm{SD}$ around the (D$^0$) and (D$^-$) resonances of sample J17 and sample H64 are depicted in Fig.\,\ref{fig1}c and Fig.\,\ref{fig1}d respectively. The red dots indicate the positions of the (D$^0$) resonance and the solid black lines crossing the red dots mark the outline of its conducting region. Sample J17 shows a first excited state at inside the conducting region (+/- 2\,mV), indicated by a solid black line, associated with the valley splitting ($\Delta$ = 2 mV) of the ground state \cite{NP}. The black dashed lines indicate V$_\mathrm{SD} = 0$. Inside the Coulomb diamond there is one electron localized on the single As donor and all the observable transport in this region finds its origin in second-order exchange processes, i.e. transport via a virtual state of the As atom. Sample J17 exhibits three clear resonances (indicated by the dashed and dashed-dotted black lines) starting from the (D$^0$) conducting region and running through the Coulomb diamond at -2,\,0 and 2\,mV. The -2\,mV and 2\,mV resonances are due to a second order transition where an electron from the source enters one valley state, an the donor-bound electron leaves from another valley state (see Fig.\,\ref{fig2}(b)). The zero bias resonance, however, is typically associated with spin Kondo effects, which happen within the same valley state. In sample H64, the pattern of the resonances looks much more complicated. We observe a resonance around 0\,mV and (interrupted) resonances that shift in V$_\mathrm{SD}$ as a function of V$_\mathrm{G}$, indicating a gradual change of the internal level spectrum as a function of V$_\mathrm{G}$. We see a large increase in conductance where one of the resonances crosses V$_\mathrm{SD} = 0$ (at V$_\mathrm{G} \sim 445 mV$, indicated by the red dashed elipsoid). Here the ground state has a full valley degeneracy, as we will show in the final paragraph. There is a similar feature in sample J17 at V$_\mathrm{G} \sim 414 mV$ in Fig.\,\ref{fig1}c (see also the red cross in Fig.\,\ref{fig1}b), although that is probably related to a nearby defect. Because of the relative simplicity of its differential conductance pattern, we will mainly use data obtained from sample J17. In order to investigate the behavior at the degeneracy point of two valley states we use sample H64. 

In the following paragraphs we investigate the second-order transport in more detail, in particular its temperature dependence, fine-structure, magnetic field dependence and dependence on $\Delta$.

We start by analyzing the temperature ($T$) dependence of sample J17. Fig.~\ref{fig2}a shows 
dI$_\mathrm{SD}$/dV$_\mathrm{SD}$ as a function of V$_\mathrm{SD}$ inside the Coulomb diamond 
(at V$_\mathrm{G}$=395 mV) for a range of temperatures. As can be readily observed from Fig.\,\ref{fig2}a, both the zero bias resonance and the two resonances at V$_\mathrm{SD}$ = +/- $\Delta$ mV are suppressed with increasing $T$. The inset of Fig.\,\ref{fig2}a shows 
the maxima $(dI/dV)_{MAX}$ of the -2 mV and 0 mV resonances as a function of $T$. We observe a logarithmic 
dependence on $T$ (a hallmark sign of Kondo correlations) at both resonances, as indicated by the red line. 
To investigate this point further we analyze another sample (H67) which has sharper resonances and of which 
more temperature-dependent data were obtained, see Fig.\,\ref{fig2}c. 
This sample also exhibits the three resonances, now at $\sim$ -1,\,0 and +1\,mV, and the same strong suppression 
by temperature. A linear background was removed for clarity. We extracted the $(dI/dV)_{MAX}$ of all three 
resonances for all temperatures and normalized them to their respective $(dI/dV)_{MAX}$ at 300 mK. 
The result is plotted in Fig.\,\ref{fig2}d. We again observe that all three peaks have the same (logarithmic) 
dependence on temperature. This dependence is described well by the following phenomenological relationship \cite{GoldhaberGordon} \begin{equation}
  \left(dI_\mathrm{SD}/dV_\mathrm{SD}\right)_{max}(T) = \left(dI_\mathrm{SD}/dV_\mathrm{SD}\right)_0 
\left( \frac{T_K^{'2}}{T^2 + T_K^{'2}} \right)^s + g_0 \label{vgl1}
\end{equation} 
where $T_K^{'}=T_K/\sqrt{2^{1/s}-1}$, $\left(dI_\mathrm{SD}/dV_\mathrm{SD}\right)_0$ is the zero-temperature 
conductance, $s$ is a constant equal to 0.22 \cite{0.22} and $g_0$ is a constant. 
Here $T_K$ is the Kondo temperature. The red curve in Fig.\,\ref{fig2}d is a fit of Eq.~(\ref{vgl1}) 
to the data. We readily observe that the data fit well and extract a $T_K$ of 2.7 K. The temperature scaling 
demonstrates that both the no valley-state flip resonance at zero bias voltage and the valley-state flip -resonance
at finite bias are due to Kondo-type processes. 

Although a few examples of finite-bias Kondo have been reported \cite{Paaske, Osorio, Roch}, the corresponding 
resonances (such as our $\pm \Delta$ resonances) are typically associated with in-elastic cotunneling. 
A finite bias between the leads breaks the coherence due to dissipative transitions in which electrons are 
transmitted from the high-potential-lead to the low-potential lead \cite{Wingreen}. These dissipative 
transitions limit the lifetime of the Kondo-type processes and, if strong enough, would only allow for 
in-elastic events. In the supporting online text we estimate the Kondo lifetime in our system and show 
it is large enough to sustain the finite-bias Kondo effects. 

The Kondo nature of the +/- $\Delta$ mV resonances points strongly towards a Valley Kondo effect \cite{Sean}, where coherent (second-order) exchange between the delocalized electrons in the contacts and the localized electron on the dopant forms a many-body singlet state that screens the {\it valley} index. Together with the more familiar spin Kondo effect, where a many-body state screens the spin index, this leads to an SU(4)-Kondo effect, where the spin and charge degree of freedom are fully entangled \cite{Pablo}. The observed scaling of the +/- $\Delta$- and zero bias -resonances in our samples by a single $T_K$ is an indication that such a fourfold degenerate SU(4)-Kondo ground state has been formed.

To investigate the Kondo nature of the transport further, we analyze the substructure of the resonances of sample J17, see Fig.\,\ref{fig2}a. The central resonance and the V$_\mathrm{SD}$ = -2\,mV each consist of three separate peaks. A similar substructure can be observed in sample H67, albeit less clear (see Fig.\,\ref{fig2}c). The substructure can be explained in the context of SU(4)-Kondo in combination with a small difference between the coupling of the ground state ($\Gamma_{GS}$)- and the first excited state ($\Gamma_{E1}$) -to the leads. It has been theoretically predicted that even a small asymmetry ($\varphi \equiv \Gamma_{E1}/\Gamma_{GS} \cong 1$) splits the Valley Kondo density of states into an SU(2)- and an SU(4) -part \cite{Lim}. This will cause both the valley-state flip- and the no valley-state flip resonances to split in three, where the middle peak is the SU(2)-part and the side-peaks are the SU(4)-parts . A more detailed description of the substructure can be found in the supporting online text. The splitting between middle and side-peaks should be roughly on the order of  T$_K$ \cite{Lim}. 
The measured splitting between the SU(2)- and SU(4) -parts equals about 0.5\,meV for sample J17 and 0.25\,meV 
for sample H67, which thus corresponds to T$_K$ $\cong$ 6\,K and T$_K$ $\cong$ 3\,K respectively, 
for the latter in line with the Kondo temperature obtained from the temperature dependence.  
We further note that dI$_\mathrm{SD}$/dV$_\mathrm{SD}$ is smaller than what we would expect for the 
Kondo conductance at T $<$ T$_K$. However, the only other study of the Kondo effect in Silicon where T$_K$ could be 
determined showed a similar magnitude of the Kondo signal \cite{SanquerKondo}. The presence of this 
substructure in both the valley-state flip-, and the no valley-state flip -Kondo resonance thus also 
points at a Valley Kondo effect.

\begin{figure}
\includegraphics[width=8cm,clip,trim=0cm 0cm 0cm 0cm]{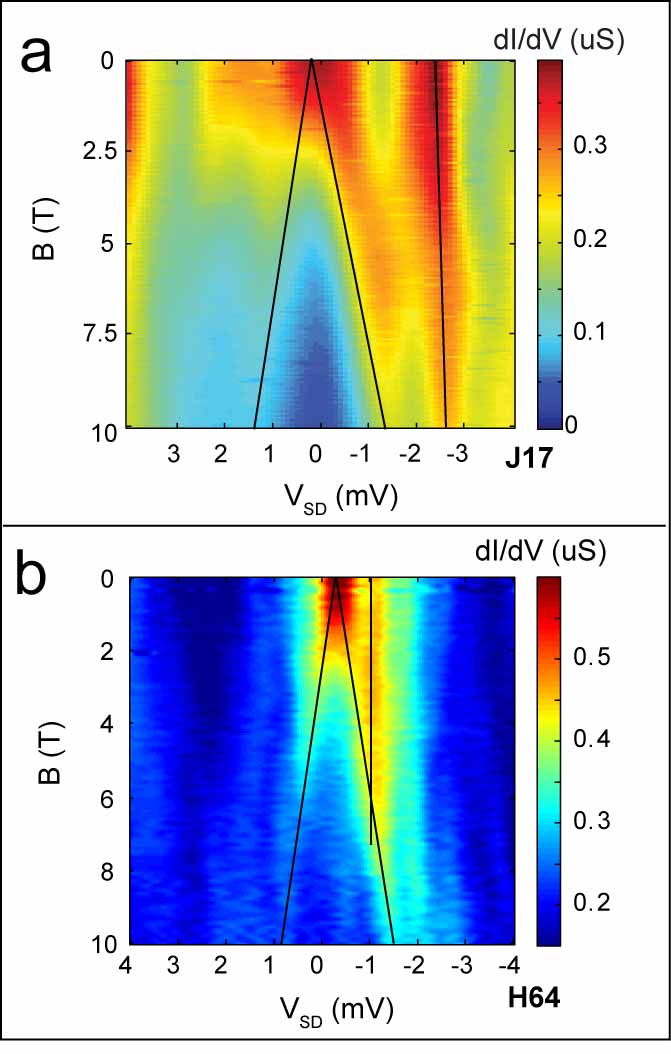}
\caption{Colormap plot of the conductance as a function of V$_\mathrm{SD}$ and $B$ of sample 
J17 at V$_G$ = 395\,mV (a) and H64 at V$_G$ = 464\,mV (b). The central Kondo resonances split in two lines which are separated by $2g^{*}\mu_{B}B$. The resonances with a valley-state flip do not seem to split in magnetic field, a feature we associate with 
the different decay-time of parallel and anti-parallel spin-configurations of the doubly-occupied virtual state 
(see text).} \label{fig3}
\end{figure}

As a third step, we turn our attention to the magnetic field ($B$) dependence of the resonances. Fig.~\ref{fig3} 
shows a colormap plot of dI$_\mathrm{SD}$/dV$_\mathrm{SD}$ for samples J17 and H64 both as a function of 
V$_\mathrm{SD}$ and $B$ at 300\,mK. The traces were again taken within the Coulomb diamond. At finite magnetic field, 
the central Kondo resonances of both devices split in two with a splitting of 2.2-2.4\,mV at $B$ = 10\,T. 
From theoretical considerations we expect the central Valley Kondo resonance to split in two by 
$\Delta_B$ = $2g^*\mu_{B}B$ if there is {\it no} mixing of valley index (this typical $2g^*\mu_{B}B$-splitting 
of the resonances is one of the hallmarks of the Kondo effect \cite{Wingreen}), and to split in three 
(each separated by $g^*\mu_{B}B$) if there {\it is} a certain degree of valley index mixing \cite{Sean}. 
Here, $g^*$ is the g-factor (1.998 for As in Si) and $\mu_B$ is the Bohr magneton. In the case of full 
mixing of valley index, the valley Kondo effect is expected to vanish and only spin Kondo will remain \cite{Lim}. 
By comparing our measured magnetic field splitting ($\Delta_B$) with $2g^*\mu_{B}B$, we find a g-factor between 
2.1 and 2.4 for all three devices. This is comparable to the result of Klein {\it et al.} who found a g-factor 
for electrons in SiGe quantum dots in the Kondo regime of around 2.2-2.3 \cite{Klein}.

The magnetic field dependence of the central resonance indicates that there is no significant mixing of valley index. 
This is an important observation as the occurrence of Valley Kondo in Si depends on the absence of mixing 
(and thus the valley index being a good quantum number in the process). The conservation of valley index 
can be attributed to the symmetry of our system. The large 2D-confinement provided by the electric field 
gives strong reason to believe that the ground- and first excited -states, $E_{GS}$ and $E_1$, consist of 
(linear combinations of) the $\textbf{k} = (0,0,\pm k_z)$ valleys (with $z$ in the electric field direction) \cite{NP, Hada}. As momentum perpendicular to the tunneling direction ($k_x$, see Fig.\,\ref{fig1}) is conserved, also valley index is conserved in tunneling \cite{Eto2}. The $\textbf{k} = (0,0,\pm k_z)$-nature of $E_{GS}$ and $E_1$ should be associated with the absence of significant exchange interaction between the two states which puts them in the non-interacting limit, and thus not in the correlated Heitler-London limit where singlets and triplets are formed. 

We further observe that the Valley Kondo resonances with a valley-state flip do not split in magnetic field, 
see Fig.\,\ref{fig3}. This behavior is seen in both samples, as indicated by the black straight solid lines, and 
is most easily observed in sample J17. These valley-state flip resonances are associated with different 
processes based on their evolution with magnetic field. The processes which involve both a valley flip and a spin flip are expected to shift to 
energies  $\pm\Delta\pm g^*\mu_{B}B$, while those without a spin-flip stay at energies $\pm\Delta$ \cite{Sean, Lim}. 
We only seem to observe the resonances at $\pm\Delta$, i.e. the valley-state flip resonances without spin flip.
In Ref \cite{Pablo}, the processes with both an orbital and a spin flip also could not be observed. 
The authors attribute this to the broadening of the orbital-flip resonances. Here, we attribute the absence 
of the processes with spin flip to the difference in life-time between the virtual valley state where 
two spins in seperate valleys are parallel ($\tau_{\uparrow \uparrow}$) and the virtual state where two 
spins in seperate valleys are anti-parallel ($\tau_{\uparrow \downarrow}$). In contrast to the latter, in the parallel spin
configuration the electron occupying the valley state with energy $E_1$, cannot decay to the other valley state at E$_{GS}$ 
due to Pauli spin blockade. It would first needs to flip its spin~\cite{lifetimes}. We have estimated $\tau_{\uparrow \uparrow}$ and $\tau_{\uparrow \downarrow}$ in our system (see supporting online text) and find that $\tau_{\uparrow \uparrow} >> h/k_{b}T_{K} > \tau_{\uparrow \downarrow}$, where $h/k_{b}T_{K}$ is the characteristic time-scale of the Kondo processes. Thus, the antiparallel spin configuration will have relaxed before it has a change to build up a Kondo resonance. Based on these lifetimes, we do not expect to observe the Kondo resonances associated with both an valley-state- and a spin -flip. 

\begin{figure}
\includegraphics[width=8cm,clip,trim=0cm 0cm 0cm 0cm]{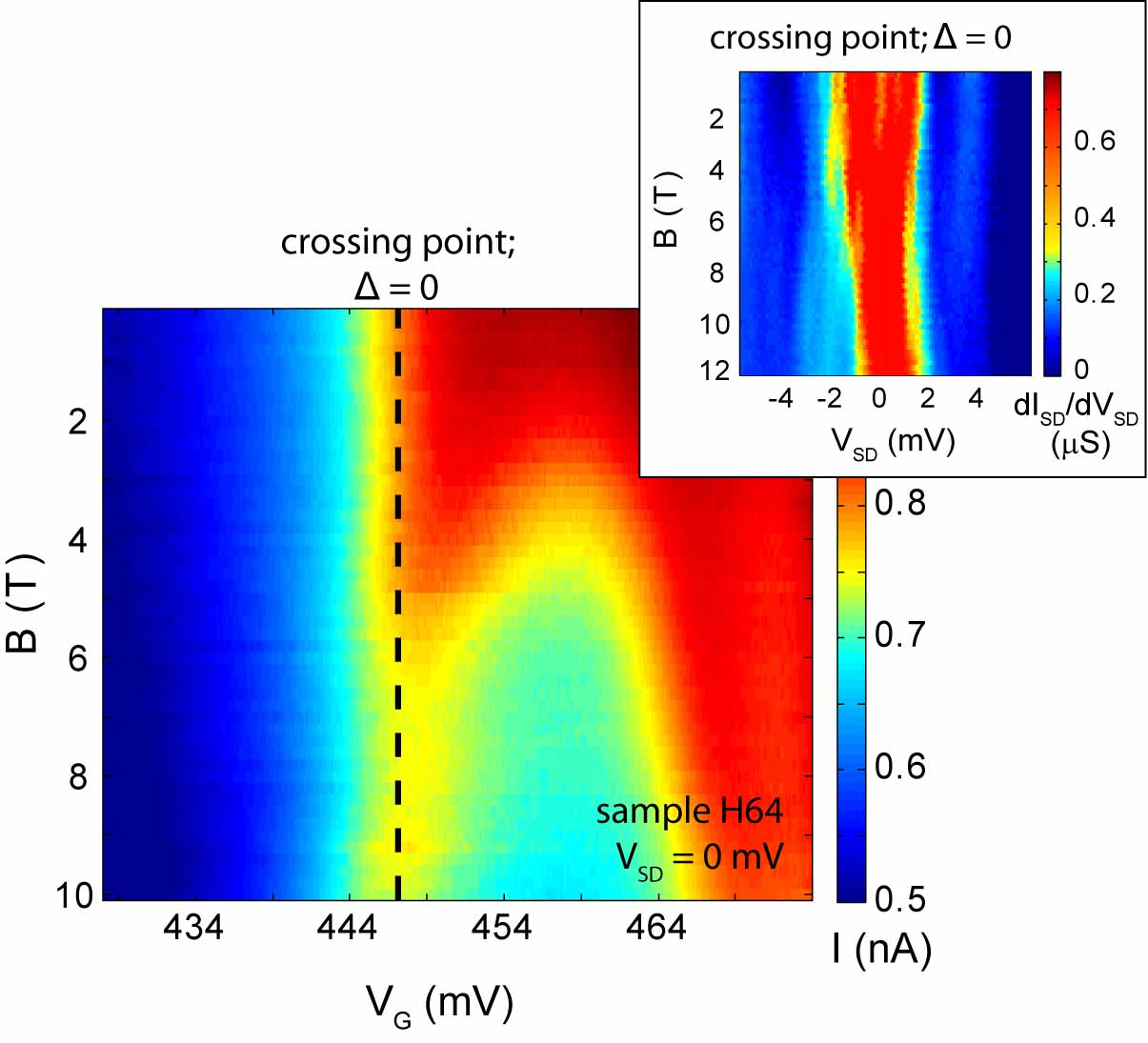}
\caption{Colormap plot of I$_{\mathrm{SD}}$ at V$_{\mathrm{SD}} = 0$ as a function of V$_{\mathrm{G}}$ and $B$. For increasing $B$, a conductance peak develops around V$_\mathrm{G} \sim 450\,mV$ at the valley degeneracy point ($\Delta$ = 0), indicated by the dashed black line. Inset: Magnetic field dependence of the valley degeneracy point. The resonance is fixed at zero bias and its magnitude does not depend on the magnetic field.} \label{fig5}
\end{figure}

Finally, we investigate the degeneracy point of valley states in the Coulomb diamond of sample H64. 
This degeneracy point is indicated in Fig.\ref{fig1}d by the red dashed ellipsoid. By means of the gate electrode, we 
can tune our system onto- or off this degeneracy point. The gate-tunability in this sample is created by a reconfiguration of the level spectrum between the D$^0$ and D$^-$- charge states, probably due to Coulomb interactions in the D$^-$- states. Figure\,\ref{fig5} shows a colormap plot of I$_\mathrm{SD}$ at V$_\mathrm{SD} = 0$ as a function of V$_\mathrm{G}$ and $B$ (at 0.3\,K). Note that we are thus looking at the 
current associated with the central Kondo resonance. At $B$ = 0, we observe an increasing I$_\mathrm{SD}$ 
for higher V$_\mathrm{G}$ as the atom's D$^-$ -level is pushed toward $E_F$. As $B$ is increased, the central Kondo 
resonance splits and moves away from V$_\mathrm{SD}$ = 0, see Fig.\,\ref{fig3}. This leads to a general decrease 
in I$_\mathrm{SD}$. However, at around V$_\mathrm{G} = 450\,mV$ a peak in I$_\mathrm{SD}$ develops, 
indicated by the dashed black line. The applied $B$-field splits off the resonances with spin-flip, but it is the valley Kondo resonance here that stays at zero bias voltage giving rise to the local current peak. The inset of Fig.\,\ref{fig5} shows the single Kondo resonance in dI$_\mathrm{SD}$/dV$_\mathrm{SD}$ as a function of V$_\mathrm{SD}$ and $B$. We observe that the magnitude of the resonance does not 
decrease significantly with magnetic field in contrast to the situation at $\Delta \neq$ 0 (Fig.\,\ref{fig3}b). 
This insensitivity of the Kondo effect to magnetic field which occurs {\it only} at $\Delta$ = 0 indicates the profound role of valley Kondo processes in our structure. It is noteworthy to mention that at this specific combination of V$_\mathrm{SD}$ and V$_\mathrm{G}$ the device can potentially work as a spin-filter \cite{Borda}. 

We acknowledge fruitful discussions with Yu.\,V. Nazarov, R. Joynt and S. Shiau. This project is supported by the Dutch Foundation for Fundamental
Research on Matter (FOM).


\newpage
\setcounter{figure}{0}

{\bf Supporting Information}

\medskip

\noindent {\bf FinFET Devices}

The FinFETs used in this study consist of a silicon nanowire connected to large contacts etched in a 60nm layer of p-type 
Silicon On Insulator. The wire is covered with a nitrided oxide (1.4\,nm equivalent SiO$_2$ thickness) and a narrow 
poly-crystalline silicon wire is deposited perpendicularly on top to form a gate on three faces. Ion implantation over the 
entire surface forms n-type degenerate source, drain, and gate electrodes while the channel protected by the gate remains 
p-type, see Fig.\,1a of the main article. The conventional operation of this n-p-n field effect transistor is to apply a positive gate voltage to create an inversion in the channel and allow a current to flow. Unintentionally, there are As donors present below the Si/SiO$_2$ interface that show up in the transport characteristics \cite{Sellier_PRL2}.

\medskip

\noindent {\bf Relation between $\Delta$ and $T_K$}

The information obtained on $T_K$ in the main article allows us to investigate the relation between the splitting ($\Delta$) of the ground ($E_{GS}$)- and first excited ($E_1$) -state and $T_K$. It is expected that $T_K$ {\it decreases} as $\Delta$ {\it increases}, since a high $\Delta$ freezes out valley-state fluctuations. The relationship between $T_K$ of an SU(4) system and $\Delta$ was calculated by Eto \cite{Eto} in a poor mans scaling approach as \begin{equation}
   \frac{k_{B}T_{K}(\Delta)}{k_{B}T_{K}(\Delta=0)} = \left(\frac{k_{B}T_{K}(\Delta=0)}{\Delta}\right)^\varphi 
  \label{vglS1}
\end{equation} where $\varphi=\Gamma_{E1}/\Gamma_{GS}$, with $\Gamma_{E1}$ and $\Gamma_{GS}$ the lifetimes of $E_1$ and $E_{GS}$ respectively. Due to the small $\Delta$ compared to the barrier height between the atom and the source/drain contact, we expect $\varphi \sim 1$. Together with $\Delta$ = 1\,meV and $T_K$ $\sim$ 2.7\,K (for sample H67) and $\Delta$ = 2\,meV and $T_K$ $\sim$ 6\,K (for sample J17), Eq.\,\ref{vglS1} yields $k_{B}T_{K}(\Delta)/k_{B}T_{K}(\Delta=0)$=$0.4$ and $k_{B}T_{K}(\Delta)/k_{B}T_{K}(\Delta=0)$=$0.3$ respectively. We can thus conclude that the relatively high $\Delta$, which separates $E_{GS}$ and $E_1$ well in energy, will certainly quench valley-state fluctuations to a certain degree but is not expected to reduce $T_K$ to a level that Valley effects become obscured.

\begin{figure}
\includegraphics[width=8cm,clip,trim=0cm 0cm 0cm 0cm]{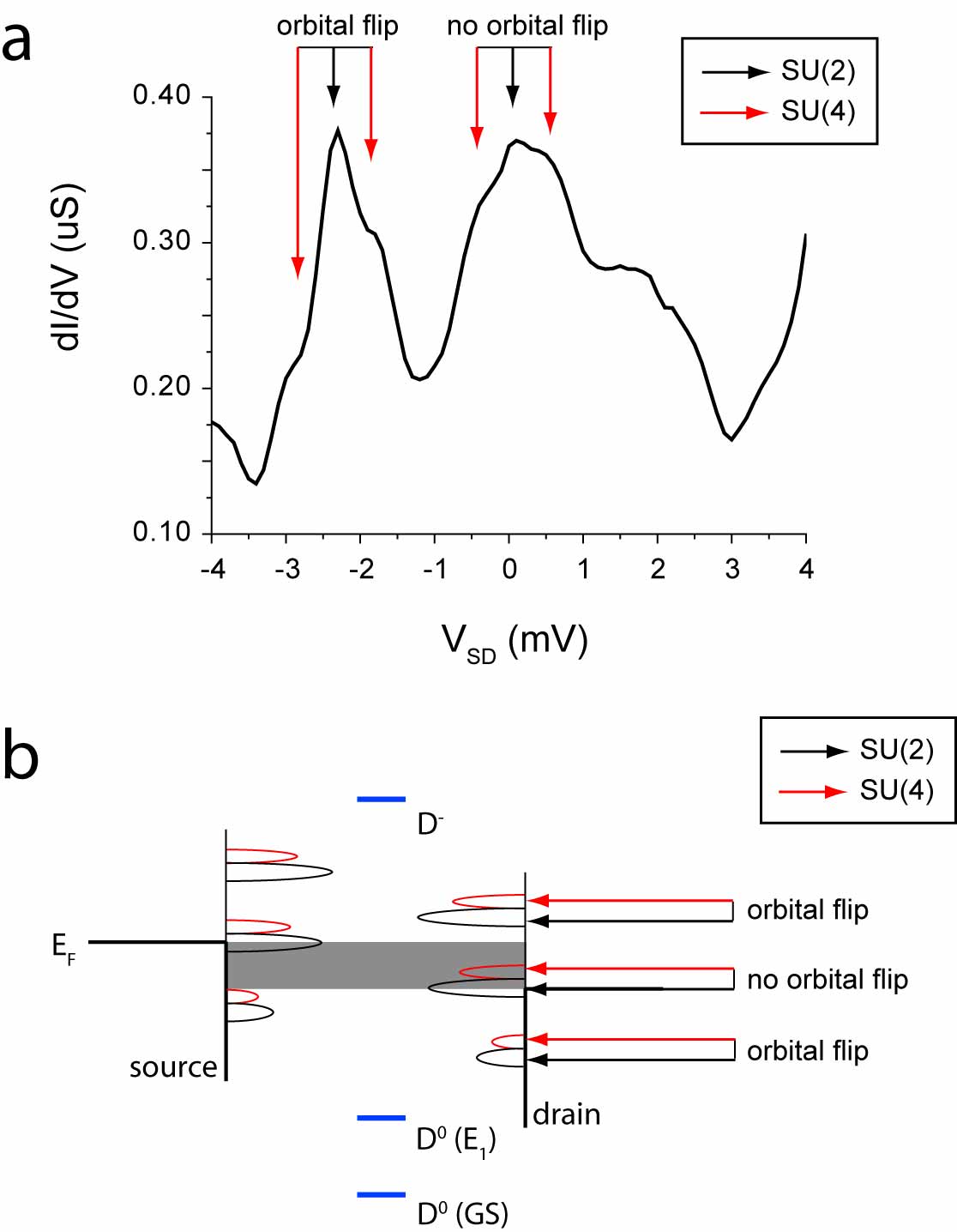}
\caption{(a) dI$_\mathrm{SD}$/dV$_\mathrm{SD}$ as a function of V$_\mathrm{SD}$ in the Kondo regime (at 395 mV$_\mathrm{G}$) of sample J17. The substructure in the Kondo resonances is the result of a small difference between $\Gamma_{E1}$ and $\Gamma_{GS}$. This splits the peaks into a (central) SU(2)-part (black arrows) and two SU(4)-peaks (red arrows).(b) Density of states in the channel as a result of $\varphi (=\Gamma_{E1}/\Gamma_{GS}) < 1$ and applied V$_\mathrm{SD}$.}\label{figS1}
\end{figure}

\medskip

\noindent {\bf Valley Kondo density of states}

Here, we explain in some more detail the relation between the density of states induced by the Kondo effects and the resulting current. The Kondo density of states (DOS) has three main peaks, see Fig.\,\ref{figS1}a. A central peak at $E_F$ = 0 due to processes without valley-state flip and two peaks at $E_F$ = $\pm \Delta$ due to processes with valley-state flip, as explained in the main text. Even a small asymmetry ($\varphi$ close to 1) will split the Valley Kondo DOS into an SU(2)- and an SU(4) -part \cite{Lim2}, indicated in Fig\,\ref{figS1}b in black and red respectively. The SU(2)-part is positioned at $E_F$ = 0 or $E_F$ = $\pm \Delta$, while the SU(4)-part will be shifted to slightly higher positive energy (on the order of $T_K$). A voltage bias applied between the source and drain leads results in the Kondo peaks to split, leaving a copy of the original structure in the DOS now at the $E_F$ of each lead, which is schematically indicated in Fig.\,\ref{figS1}b by a separate DOS associated with each contact. The current density depends directly on the density of states present {\it within} the bias window defined by source/drain (indicated by the gray area in Fig\,\ref{figS1}b) \cite{Hershfield}. The splitting between SU(2)- and SU(4) -processes will thus lead to a three-peak structure as a function of V$_\mathrm{SD}$. 

Figure.\,\ref{figS1}a has a few more noteworthy features. The zero-bias resonance is not positioned exactly at V$_\mathrm{SD} = 0$, as can also be observed in the transport data (Fig\,1c of the main article) where it is a few hundred\,$\mu$eV above the Fermi energy near the D$^0$ charge state and a few hundred\,$\mu$eV below the Fermi energy near the D$^-$ charge state. This feature is also known to arise in the Kondo strong coupling limit \cite{Horvatic, Bichler}. We further observe that the resonances at V$_\mathrm{SD}$ = +/- 2 mV differ substantially in magnitude. This asymmetry between the two side-peaks can actually be expected from SU(4) Kondo systems where $\Delta$ is of the same order as (but of course always smaller than) the energy spacing between $E_{GS}$ and $E_F$, see Ref\,\cite{Inoshita2}. The +2\,mV resonance is not large enough to clearly distinguish any substructure.

\medskip

\noindent {\bf Kondo Lifetime}

\begin{figure}
\includegraphics[width=8cm,clip,trim=0cm 0cm 0cm 0cm]{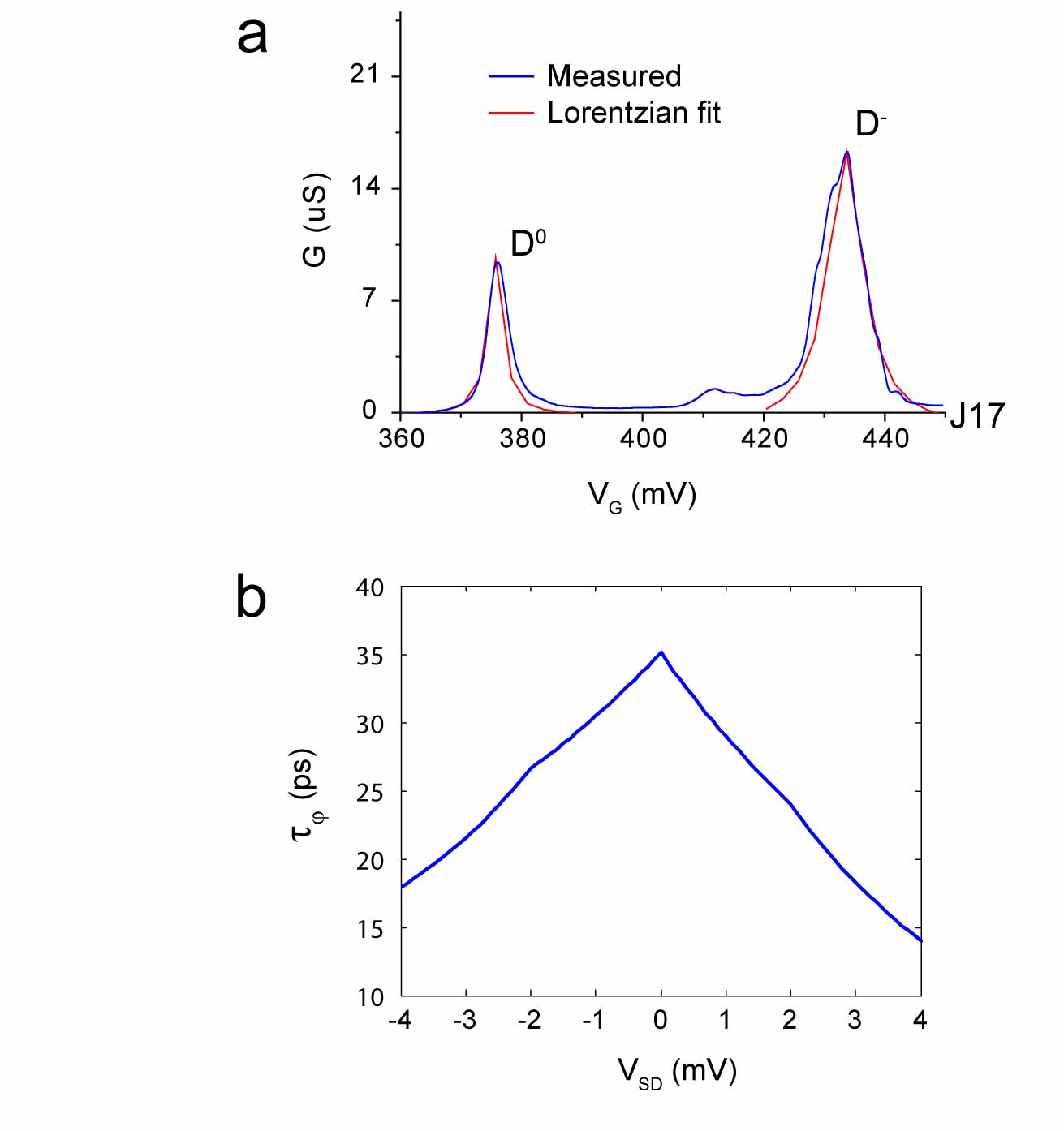}
\caption{(a) Differential conductance of sample J17 at V$\mathrm{_{SD}} = 0$ as a function of gate voltage (blue trace). The one-electron and two-electron charge states are indicated by D$^0$ and D$^-$ respectively. The red traces are Lorentzian fits to the left side of the D$^0$ and right side of D$^-$ state to extract the FWHM (b) Calculated Kondo lifetime ($\tau_{\varphi}$) versus V$_\mathrm{SD}$ for sample J17 at V$_\mathrm{G}$ = 395\,mV.}\label{figS2}
\end{figure}

We associate the valley-state flip resonances with (coherent) Kondo-type processes that occur at {\it finite} bias. However, a finite bias between the leads should break the coherence due to dissipative transitions in which electrons are transmitted from the high-potential-lead to the low-potential lead. These dissipative transitions limit the lifetime of the Kondo-type processes. Only if this so-called Kondo lifetime ($\tau_{\varphi}$) is larger than the characteristic timescale of the Kondo processes (given by $h/k_{b}T_K$), finite bias Kondo can occur. If $\tau_{\varphi} << h/k_{b}T_K$, any transport through the atom where the localized electron flips its valley-state would be necessarily caused by inelastic co-tunneling events. 

Here, we estimate the Kondo life-time of sample J17 based on a Fermi's Golden Rule approach by Wingreen and Meir \cite{Wingreen2}. Following their derivation we obtain for the lifetime $\tau_{\phi}$ of the ground (a) and first excited state (b) of the atom 
\begin{widetext}
\begin{equation}
   \frac{1}{\tau_{\phi ,j}} = \frac{1}{2 \pi \hbar} \sum_{\stackrel{A = L,R}{j'=a,b}}{\Gamma^{A}_{j} \Gamma^{B}_{j'}\Theta\left(\mu_B - \mu_A + \epsilon_j - \epsilon_j' \right) \frac{\mu_B - \mu_A + \epsilon_j - \epsilon_j'}{\left(\mu_A - \epsilon_j\right)\left( \mu_B - \epsilon_j'\right)}}
  \label{vglS3}
\end{equation}
\end{widetext}
where $j = a,b$, $\mu_A$ and $\mu_B$ are the Fermi energies of respectively the source and drain electrodes, $\epsilon_a$ and $\epsilon_b$ are the energy of the ground and first excited state and $\Gamma_L$ and $\Gamma_R$ are the tunnel rates to the source and drain respectively. We can find $\Gamma_L$ and $\Gamma_R$ of the ground state by analyzing the FWHMs and peak heights of the direct transport zero bias resonances. 

The total tunnel rate to the leads $\Gamma =\Gamma_L+ \Gamma_R$ follows from the FWHM of the D$^0$ and D$^-$ charge resonances, see fig.\,\ref{figS2}a. We fit these resonances by a Lorentzian peak shape (we only fit the left side of the D$^0$- and right side of D$^-$ -resonances as both peaks are slightly asymmetric due to the Kondo effect), indicated by the red traces. The FWHM of the resonances is a result of the life-time broadening $\Gamma$ convoluted with the gaussian Fermi distribution of the electrons in the lead, with a FWHM of $3.5 k_{b} T$. We extract $\Gamma$ from the fitted FWHM by deconvolution \cite{Wilkinson}, and find $\Gamma^{D^0}$ = 1.6 meV, $\Gamma^{D^-}$ = 4.1 meV.

We obtain the ratio between $\Gamma_L$ and $\Gamma_R$ from the resonance peak conductions. We can readily observe that the system is in the coherent transport limit of $h\Gamma >> k_{b}T$ as $FWHM \cdot e >> k_{b} T$ \cite{Beenakker} where $\Gamma$ is the tunnel rate to the leads, see Fig.\,\ref{figS2}. Here the conductance can be written as \cite{Buttiker}
\begin{equation}
   G =  \frac{4 e^2}{\hbar} \frac{\Gamma_L \Gamma_R}{\left( \Gamma_L + \Gamma_R \right)^2} 
  \label{vglS2}
\end{equation}
where $\Gamma_L$ and $\Gamma_R$ are the tunnel rates to the left and right lead respectively. The resonance maximums are obtained from Fig\,\ref{figS2}a and combined with Eq.\,\ref{vglS2} and $\Gamma^{D^0}$ = 1.6 meV, $\Gamma^{D^-}$ = 4.1 meV we find $\Gamma^{D^0}_L$ = 1.4 meV, $\Gamma^{D^0}_R$ = 0.2 meV, $\Gamma^{D^-}_L$ = 3.1 meV and $\Gamma^{D^-}_R$ = 0.9 meV.

We assume the coupling of the first excited state to the left and reight leads is equal to that of the ground state. This assumption is justified by the smallness of the valley splitting ($\Delta$) compared to the barrier heights and the comparable symmetry of the ground and first excited states.

As such, we have calculated $\tau_{\varphi}$ for sample J17 as a function of V$\mathrm{_{SD}}$ at V$\mathrm{_{G}}$ = 395\,mV, see Fig\,\ref{figS2}a. The values for $\mu_A$, $\mu_B$, $\epsilon_j$ and $\epsilon_{j'}$ were extracted from the stability diagram. The figure thus shows $\tau_{\varphi}$ for the measurements in Fig.\,2(a) and Fig.\,3(a) of the main article. The estimated $\tau_{\varphi}$ at +/- 2mV is 25\,ps. It should be noted that Eq.\,\ref{vglS3} captures only second-order fluctuations and ignores higher-order corrections to $\tau_{\varphi}$. We have calculated the fourth-order corrections on $\tau_{\varphi}$ and found them to be insignificant compared to the contribution of second-order fluctuations. The estimated $\tau_{\varphi}$ is significantly larger than $\hbar/k_{b}T_K$, which is about 8 ps at $T_K$ = 6\,K, showing that finite-bias Kondo processes are very well possible in this system and we do not expect them to be significantly decohered by in-elastic co-tunneling processes.

\begin{figure}
\includegraphics[width=8cm,clip,trim=0cm 0cm 0cm 0cm]{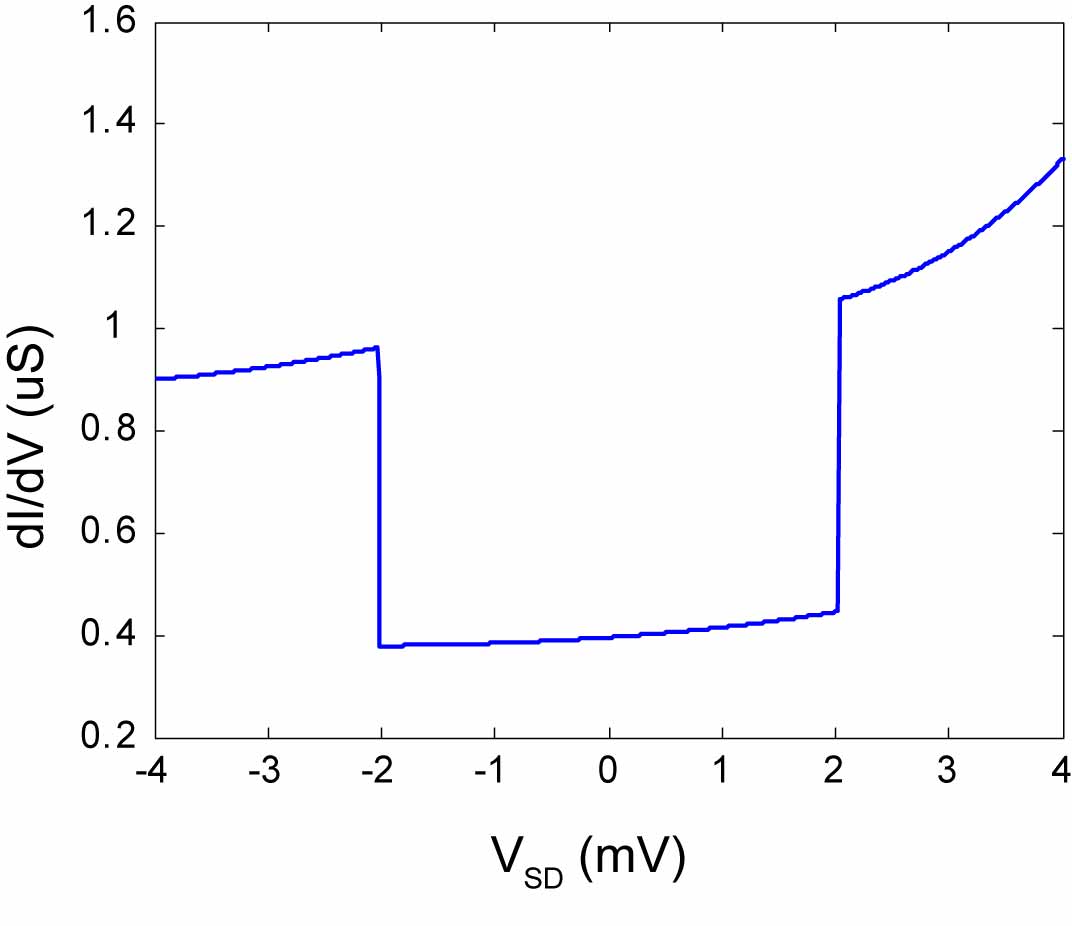}
\caption{Conductance of sample J17 as a function of V$\mathrm{_{SD}}$ again at V$\mathrm{_{G}}$ = 395\,mV if it would be in the cotunneling regime, so if ($\tau_{\varphi} >> \hbar/k_{b}T_K$). The graph can be compared directly with the measurements in Fig.\,2a and Fig.\,3 of the main article} \label{figS3}
\end{figure}

Figure\,\ref{figS3} shows a calculation of the conductance of sample J17 as a function of V$\mathrm{_{SD}}$ again at V$\mathrm{_{G}}$ = 395\,mV if it would be in the cotunneling regime ($\tau_{\varphi} >> \hbar/k_{b}T_K$). Here we follow the derivation by Wegewijs and Nazarov \cite{Wegewijs}. There are steps at $\pm$ 2\,mV where the in-elastic cotunneling via the first excited state sets in. As we can readily observe, the calculation suggests that no significant peak structure occurs on top of the steps at $\pm$ 2\,mV as a result of change of occupation in the states involved. This provides another clue that the $\pm$ $\Delta$ resonances are not caused by in-elastic co-tunneling.

\medskip

\noindent {\bf Spin lifetimes}

\begin{figure}
\includegraphics[width=8cm,clip,trim=0cm 0cm 0cm 0cm]{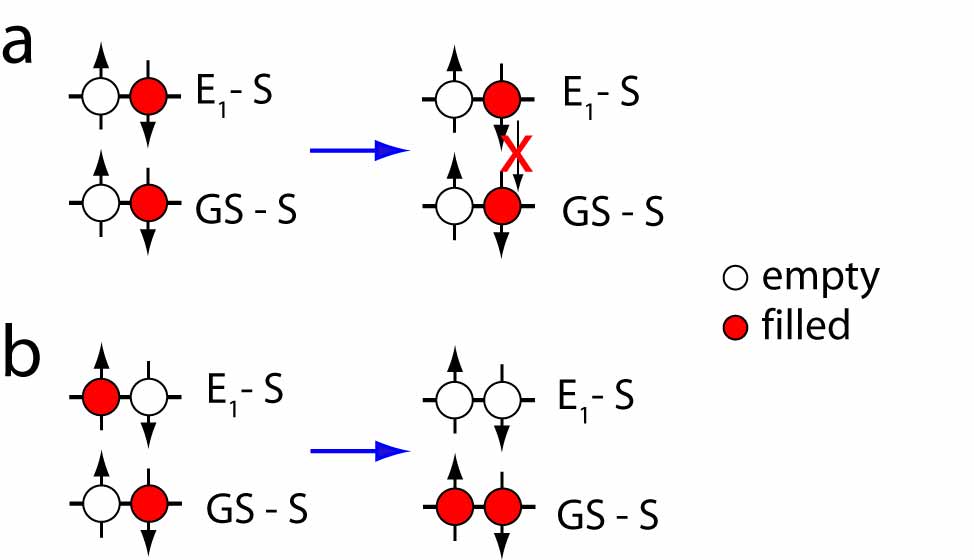}
\caption{Schematic depiction of the virtual two electron state during the Kondo process. The parallel spin configuration (a) first needs to spin-flip to the anti-parallel configuration (b) before a recombination can take place. When two electron occupy either the first excited state $E_1$ or ground state $GS$ they necessarily have to form singlets. However, when two electrons are on different valley states, due to the absence of exchange interaction, no singlet-triplet formation takes place and the system is in the non-interacting limit.} \label{figS4}
\end{figure}

We estimate the decay time of the anti-parallel spin configuration ($\tau_{\uparrow \downarrow}$) from the lifetime of the $1s(E)$, $1s(T_2)$ excited states of Arsenic donors. These lifetimes were experimentally determined for bulk As donors to be around 10-100\,ps \cite{Castner}, but are expected to be much smaller in our system due to the close vicinity of the Si/SiO$_2$ interface. The parallel spin configurations (with life-time $\tau_{\uparrow \uparrow}$) first needs to spin-flip to the anti-parallel configuration before a recombination can take place, see Fig.\,\ref{figS4}. A safe lower bound for the time-scale for such a spin-flip event is estimated from the $T_1$ of confined electrons in the channel of Si FETs, which has been experimentally determined to be 460\,ns \cite{Lyon}. The decay-times of both spin-configurations should be compared with the characteristic time-scale of the Kondo processes, which follows directly from the measured Kondo temperature as $h/k_{b}T_{K} = 8 ps$. We can thus expect that $\tau_{\uparrow \uparrow} >> h/k_{b}T_{K} > \tau_{\uparrow \downarrow}$, which explains why only the processes without spin-flip are observed in Fig.\,3 of the main article.

\begin{thebibliography}{30}

\bibitem{JunKondo}
Kondo, J., Resistance Minimum in Dilute Magnetic Alloys, {\it Prog. Theor. Phys.} {\bf 32} 37-49 (1964)

\bibitem{Hewson}
Hewson, A.C., {\it The Kondo Problem to Heavy Fermions} (Cambridge Univ. Press, Cambridge, 1993).

\bibitem{WingreenOverview}
Wingreen N.S., The Kondo effect in novel systems, {\it Mat. Science Eng. B} {\bf 84} 22–25 (2001) and references therein.

\bibitem{Cox}
Cox, D.L., Zawadowski, A., Exotic Kondo effects in metals: magnetic ions in a crystalline electric field and tunneling centers, {\it Adv. Phys.} {\bf 47}, 599-942 (1998)

\bibitem{Inoshita}
Inoshita, T., Shimizu, A., Kuramoto, Y., Sakaki, H., Correlated electron transport through a quantum dot: the multiple-level effect. {\it Phys. Rev. B} {\bf 48}, 14725 - 14728 (1993)

\bibitem{Borda}
Borda, L. Zar$\mathrm{\acute{a}}$nd, G., Hofstetter,W., Halperin, B.I. and von Delft, J., SU(4) Fermi Liquid State and Spin Filtering in a Double Quantum Dot System, {\it Phys. Rev. Lett.} {\bf 90}, 026602 (2003) 

\bibitem{Zarand}
Zar$\mathrm{\acute{a}}$nd, G., Orbital fluctuations and strong correlations in quantum dots, Philosophical Magazine, 86, 2043 - 2072 (2006)

\bibitem{Pablo}
Jarillo-Herrero, P., Kong, J., van der Zant H.S.J., Dekker, C., Kouwenhoven, L.P., De Franceschi, S., Orbital Kondo effect in carbon nanotubes, {\it Nature} {\bf 434}, 484 (2005)

\bibitem{Koiller}
Martins, A.S., Capaz, R.B. and Koiller, B., Electric-field control and adiabatic evolution of shallow donor impurities in silicon, {\it Phys. Rev. B} {\bf 69}, 085320 (2004)

\bibitem{NP}
Lansbergen, G.P. {\it et al.}, Gate induced quantum confinement transition of a single dopant atom in a Si FinFET, {\it Nature Physics} {\bf 4}, 656 (2008)

\bibitem{Rokhinson}
Rokhinson, L.P., Guo, L.J., Chou, S.Y., Tsui, D.C., Kondo-like zero-bias anomaly in electronic transport through an ultrasmall Si quantum dot, {\it Phys. Rev. B} {\bf 60}, R16319 - R16321 (1999)

\bibitem{SanquerKondo}
Specht, M., Sanquer, M., Deleonibus, S., Gullegan G., Signature of Kondo effect in silicon quantum dots, {\it Eur. Phys. J. B} {\bf 26}, 503-508 (2002)

\bibitem{Klein}
Klein, L.J., Savage, D.E., Eriksson, M.A., Coulomb blockade and Kondo effect in a few-electron silicon/silicon-germanium quantum dot, {\it Appl. Phys. Lett.} {\bf 90}, 033103 (2007) 

\bibitem{Sean}
Shiau, S., Chutia, S. and Joynt, R., Valley Kondo effect in silicon quantum dots, {\it Phys. Rev. B} {\bf 75}, 195345 (2007)

\bibitem{Roch}
Roch, N., Florens, S., Bouchiat, V., Wernsdirfer, W., Balestro, F., Quantum phase transistion in a single molecule quantum dot, {\it Nature} {\bf 453}, 633 (2008)

\bibitem{Sellier_PRL}
Sellier, H. {\it et~al.}, Transport Spectroscopy of a Single Dopant in a Gated Silicon Nanowire, {\it Phys. Rev. Lett.} {\bf 97}, 206805 (2006)

\bibitem{Calvet}
Calvet, L.E., Wheeler, R.G. and Reed, M.A., Observation of the Linear Stark Effect in a Single Acceptor in Si, {\it Phys. Rev. Lett.} {\bf 98}, 096805 (2007)

\bibitem{Sanquer}
Hofheinz, M. {\it et al.}, Individual charge traps in silicon nanowires, {\it Eur. Phys. J. B} {\bf 54}, 299–307 (2006)

\bibitem{condmatgrenoble}
Pierre, M., Hofheinz, M., Jehl, X., Sanquer, M., Molas, G.,Vinet, M., Deleonibus S., Offset charges acting as excited states in quantum dots spectroscopy, {\it 	Eur. Phys. J. B} {\bf 70}, 475-481 (2009)

\bibitem{GoldhaberGordon}
Goldhaber-Gordon, D., Göres, J., Kastner, M.A., Shtrikman, H., Mahalu, D., Meirav, U., From the Kondo Regime to the Mixed-Valence Regime in a Single-Electron Transistor, {\it Phys. Rev. Lett.} {\bf 81}, 5225 (1998)

\bibitem{0.22}
Although the value of $s =0.22$ stems from SU(2) spin Kondo processes, it is valid for SU(4)-Kondo systems as 
well \cite{Lim, Pablo}.

\bibitem{Paaske} 
Paaske, J., Rosch, A., W$\mathrm{\ddot{o}}$lfle, P., Mason, N., Marcus, C.M., Nyg$\mathrm{\dot{a}}$rd, Non-equilibrium singlet-triplet Kondo effect in carbon nanotubes, {\it Nature Physics} {\bf 2}, 460 (2006)

\bibitem{Osorio}
Osorio, E.A. {\it et al.}, Electronic Excitations of a Single Molecule Contacted in a Three-Terminal Configuration, {\it Nanoletters} {\bf 7}, 3336-3342 (2007)

\bibitem{Wingreen}
Meir, Y., Wingreen, N.S., Lee, P.A., Low-Temperature Transport Through a Quantum Dot: The Anderson Model Out of Equilibrium, {\it Phys. Rev. Lett.} {\bf 70}, 2601 (1993)

\bibitem{Lim}
Lim, J.S., Choi, M-S, Choi, M.Y., L$\mathrm{\acute{o}}$pez, R., Aguado, R., Kondo effects in carbon nanotubes: From SU(4) to SU(2) symmetry, {\it Phys. Rev. B} {\bf 74}, 205119 (2006)

\bibitem{Hada}
Hada, Y., Eto, M., Electronic states in silicon quantum dots: Multivalley artificial atoms, {\it Phys. Rev. B} {\bf 68}, 155322 (2003)

\bibitem{Eto2}
Eto, M., Hada, Y., Kondo Effect in Silicon Quantum Dots with Valley Degeneracy, {\it AIP Conf. Proc.} {\bf 850}, 1382-1383 (2006)  

\bibitem{lifetimes}
A comparable process in the direct transport through Si/SiGe double dots (Lifetime Enhanced Transport) 
has been recently proposed \cite{Eriksson}. 

\bibitem{Eriksson}
Shaji, N. {\it et. al.}, Spin blockade and lifetime-enhanced transport in a few-electron Si/SiGe double quantum dot, {\it Nature Physics} {\bf 4}, 540 (2008)

\end{thebibliography}

\begin{thebibliography}{30}

\bibitem{Sellier_PRL2}
Sellier, H. {\it et~al.}, Transport Spectroscopy of a Single Dopant in a Gated Silicon Nanowire, {\it Phys. Rev. Lett.} {\bf 97}, 206805 (2006)

\bibitem{Eto}
Eto, M., Enhancement of Kondo effect in multilevel quantum dots {\it J. Phys. Soc. Japan} {\bf 74}, 95-102 (2005)

\bibitem{Lim2}
Lim, J.S., Choi, M-S, Choi, M.Y., L$\mathrm{\acute{o}}$pez, R., Aguado, R., Kondo effects in carbon nanotubes: From SU(4) to SU(2) symmetry, {\it Phys. Rev. B} {\bf 74}, 205119 (2006)

\bibitem{Hershfield}
Hershfield, S., Davies, J.H., Wilkins, J.W., Probing the Kondo Resonance by Resonant Tunneling through an Anderson Impurity, {\it Phys. Rev. Lett} {\bf 67}, 3720 (1991)

\bibitem{Horvatic}
Horvati$\mathrm{\acute{c}}$, B., $\mathrm{\breve{S}}$ok$\mathrm{\breve{c}}$evi$\mathrm{\acute{c}}$, D., Zlati$\mathrm{\acute{c}}$, V., Finite-temperature spectral density for the Anderson model, {\it Phys. Rev. B} {\bf 36}, 675-683 (1987)
Horvati

\bibitem{Bichler}
Simmel, F., Blick, R.H., Kotthaus, J.P.,W egscheider, W., Bichler, M., Anomalous Kondo Effect in a Quantum Dot at Nonzero Bias, {\it Phys. Rev. Lett} {\bf 83}, 804 (1999)

\bibitem{Inoshita2}
Inoshita, T., Shimizu, A., Kuramoto, Y., Sakaki, H., Correlated electron transport through a quantum dot: the multiple-level effect. {\it Phys. Rev. B} {\bf 48}, 14725 - 14728 (1993)

\bibitem{Wingreen2}
Wingreen, N.S. and Meir, Y., Anderson model out of equilibrium: Noncrossing-approximation approach to transport through a quantum dot, {\it Phys. Rev. B} {\bf 49}, 11040-11052 (1994)

\bibitem{Wilkinson}
Wilkinson, D.H., Breit-Wigners viewed through Gaussians, {\it Nuclear instruments and methods} {\bf 95}, 259-264 (1971)

\bibitem{Beenakker}
Beenakker, C.W.J., Theory of Coulomb-blockade oscillations in the conductance of a quantum dot, {\it Phys. Rev. B} {\bf 44}, 1646-1656 (1991)

\bibitem{Buttiker}
B$\mathrm{\ddot{u}}$ttiker, M., Coherent and sequantial tunneling in series barriers, {\it IBM J. Res. Develop.} {\bf 32}, 63-75 (1988)

\bibitem{Wegewijs}
Wegewijs, M.R., Nazarov, Yu.V., In-elastic cotunneling through an excited state of a quantum dot, {\it arXiv:0103579v2 [cond-mat.mes-hall]} (2001)

\bibitem{Castner}
Castner, T.G., Raman Spin-Lattice Relaxtion of Shallow Donors in Silicon, {\it Physical Review} {\bf 130}, 58-77 (1963)

\bibitem{Lyon}
Tyryshkin, A.M. {\it et. al.}, Electron spin coherence in Si, {\it Physica E} {\bf 35}, 257-263 (2006)

\end{thebibliography}
\end{document}